\documentclass[twocolumn,twocolappendix]{aastex631}
\shorttitle{An eROSITA Survey of Virgo and Fornax AGNs}

\graphicspath{{./}{figures/}}
\usepackage{graphicx}

\begin{document}

\title{An X-ray Census of Active Galactic Nuclei in the Virgo and Fornax Clusters of Galaxies with SRG/eROSITA}

%\correspondingauthor{August Muench}
%\email{greg.schwarz@aas.org, gus.muench@aas.org}

\author[0000-0001-9062-8309]{Meicun Hou}
\affiliation{Kavli Institute for Astronomy and Astrophysics, Peking University, Beijing 100871, China}
\email{houmc@pku.edu.cn} 
\author[0009-0009-9972-0756]{Zhensong Hu}
\affiliation{School of Astronomy and Space Science, Nanjing University, Nanjing $210023$, China}
\affiliation{Key Laboratory of Modern Astronomy and Astrophysics (Nanjing University), Ministry of Education, Nanjing 210023, China}
\email{huzhensong@smail.nju.edu.cn}
\author[0000-0003-0355-6437]{Zhiyuan Li}
\affiliation{School of Astronomy and Space Science, Nanjing University, Nanjing $210023$, China}
\affiliation{Key Laboratory of Modern Astronomy and Astrophysics (Nanjing University), Ministry of Education, Nanjing 210023, China}
\email{lizy@nju.edu.cn}

\begin{abstract}
We present a uniform and sensitive  X-ray census of active galactic nuclei (AGNs) in the two nearest galaxy clusters, Virgo and Fornax, utilizing the newly released X-ray source catalogs from the first all-sky scan of SRG/eROSITA. A total of 50 and 10 X-ray sources are found positionally coincident with the nuclei of member galaxies in Virgo and Fornax, respectively, down to a 0.2--2.3 keV luminosity of $\sim10^{39}\rm~erg~s^{-1}$ and reaching out to a projected distance well beyond the virial radius of both clusters. 
The majority of the nuclear X-ray sources are newly identified. 
There is weak evidence that the nuclear X-ray sources are preferentially found in late-type hosts. Several hosts are dwarf galaxies with a stellar mass below $\sim10^{9}\rm~M_\odot$.
We find that contamination by non-nuclear X-ray emission can be neglected in most cases, indicating the dominance of a genuine AGN. In the meantime, no nuclear X-ray source exhibits a luminosity higher than a few times $10^{41}\rm~erg~s^{-1}$, which might be owing to a steep intrinsic luminosity function.
The X-ray AGN occupation rate is only $\sim$3\% in both clusters, apparently much lower than that in field galaxies inferred from previous X-ray studies. Both aspects suggest that the cluster environment effectively suppresses AGN activity. 
The findings of this census have important implications on the interplay between  galaxies and their central massive black holes in cluster environments.
\end{abstract}


\section{Introduction} \label{sec:intro}
Active galactic nuclei (AGNs), the manifestation of accreting super-massive black holes (SMBHs) at the center of most galaxies \citep{2017A&ARv..25....2P}, play an indispensable and increasingly crucial role in our understanding of the formation and growth of these exotic objects, as well as their interplay and co-evolution with the host galaxies \citep{2012ARA&A..50..455F,2013ARA&A..51..511K}.

From the very beginning \citep{1963Natur.197.1040S}, AGN studies, in particular the demography of AGNs, have benefitted from two lines of research: towards greater distances and towards larger samples.
The recently launched James Webb Space Telescope with its unprecedented sensitivity in the infrared window is now unveiling some of the first-generation AGNs at the early universe \citep{2023ApJ...955L..24G}. 
On the other hand, wide-field, even all-sky, surveys of AGNs in the low-redshift universe continue to offer new insights and surprises by
revealing a vastly large number of AGNs otherwise lurking at the center of normal and dwarf galaxies. 
Most of these low-redshift AGNs turn out to have low luminosities \citep{2008ARA&A..46..475H}, which trace weakly accreting SMBHs fed by a radiatively inefficient, hot accretion flow \citep{2014ARA&A..52..529Y}. 
Growing attention has been paid by recent observational and theoretical studies about the importance of such weakly accreting SMBHs in our understanding of the SMBH-host galaxy co-evolution \citep[e.g.][]{2017MNRAS.465.3291W,2019ApJ...885...16Y,2021NatAs...5..928S}.
Alternatively, a small fraction of these low-luminosity AGNs could be smaller black holes accreting at a rate closer to the Eddington limit. Such dwarf MBHs or intermediate-mass black holes (IMBHs) are the long-sought population that bridges the stellar-mass black holes and SMBHs \citep{2020ARA&A..58..257G}.

X-rays, which trace the hot plasma in and around the accretion flow onto the SMBH, provide arguably the most direct evidence for AGNs. This has been demonstrated, in particular, by the {\it Chandra} Deep Field surveys of distant AGNs up to a redshift $z \gtrsim 6$ \citep{2002ApJS..139..369G,2017ApJS..228....2L,2017NewAR..79...59X}. 
The superb sensitivity and angular resolution of {\it Chandra} has also enabled routine detections of nuclear X-ray sources down to the Eddington limit of solar-mass objects ($\sim 10^{38}\rm~erg~s^{-1}$) in nearby galaxies \citep[e.g.][]{2008ApJ...680..154G,2017ApJ...835..223S}, enabling the study of specific AGNs as well as AGN demography across diverse galactic environments. 

The environment is believed to play a crucial role in galaxy evolution. 
In particular, in galaxy clusters, where individual galaxies travel at a typical velocity of $\sim10^3\rm~km~s^{-1}$ within the hot intra-cluster medium (ICM) and frequently interact with other galaxies, environmental effects such as ram pressure stripping \citep{2022A&ARv..30....3B} and tidal interaction continue to shape the member galaxies and their interstellar medium (ISM). It is generally expected that the long-term outcome of such effects would be to reduce the amount of the ISM, quench star formation and suppress the SMBH accretion. 

Observational evidence for starving SMBHs in the nearest galaxy cluster, Virgo, was obtained by the AGN Multi-wavelength Survey of Early-Type Galaxies in the Virgo Cluster (AMUSE-Virgo; \citealp{2008ApJ...680..154G,2010ApJ...714...25G}), which probed nuclear X-ray sources in 100 early-type galaxies (ETGs; including elliptical, lenticular, and dwarf elliptical/lenticular), finding an occupation fraction of $24\%-34\%$ down to an X-ray luminosity of few times $10^{38}\rm~erg~s^{-1}$.
A similar occupation fraction of X-ray AGNs is found for a sample of ETGs in the second nearest cluster, Fornax \citep{2019ApJ...874...77L}, and the third 
nearest cluster, Antlia \citep{2023ApJ...956..104H}, down to a comparable detection limit. 
On the other hand, among the AMUSE-Field sample of 103 nearby field and group ETGs, collectively designed as a `field' comparison for AMUSE-Virgo, \citet{2012ApJ...747...57M} reported a higher X-ray AGN occupation fraction of $45\%\pm7\%$ and a higher nuclear X-ray luminosity at a given black hole mass than the Virgo ETG sample.
\citet{2022MNRAS.512.3284S} and \citet{2021ApJ...923..246G} have gathered a sample of 75 late-type galaxies (LTGs, including normal disks and dwarf irregulars) in Virgo, finding more than half hosting a nuclear X-ray source, again down to a detection limit of few times $10^{38}\rm~erg~s^{-1}$.
 
These studies, all based on sensitive {\it Chandra} observations, provide an important insight on how the AGN activity may be related to the intrinsic properties of host galaxies continuously being modified by the environmental effects. 
Nevertheless, the sample galaxies covered by these programs are just a small fraction of the identified member galaxies in the respective clusters, and perhaps more unsatisfactorily, are biased to large galaxies and/or galaxies in the cluster core due to the design of the individual programs. 

Thanks to the advent of the SRG/eROSITA all-sky survey \citep{2021A&A...656A.132S}, a uniform and sensitive soft-X-ray census of AGNs in the two nearest galaxy clusters is within reach for the first time.  
Although the point source sensitivity of eROSITA is no better than that of {\it Chandra} due to its moderate point-spread function (with a half-energy width [HEW] $\sim 30\arcsec$; \citealp{2024A&A...682A..34M}), a sensitivity limit down to $\sim10^{39}\rm~erg~s^{-1}$ is readily achievable, sufficient to detect low-luminosity AGNs expected to be prevalent in these two clusters.
We take up such a task in this work, utilizing the freshly released eROSITA data from its first all-sky scan (eRASS1).

The remainder of this paper is structured as follows. 
The catalogues of Virgo and Fornax galaxies are described in Section~\ref{sec:catalog}. 
The identification of nuclear X-ray sources from eRASS1 are described in Section \ref{sec:nucleus}. Statistical analyses of the nuclear X-ray sources are presented in Section \ref{sec:stat}, focusing on the X-ray AGN occupation fraction as a function of various galaxy properties. 
A brief summary and some immediate implications of our findings are addressed in Section~\ref{sec:sum}.
%Section \ref{sec:summary} summarizes and concludes the AMUSE-Antlia nuclear X-ray findings. % Section \ref{sec:discussion} discussed the possible environmental influences on SMBH activity

\begin{figure*}[ht!]
\epsscale{0.8}
%\plotone{AN2S10_DSS_AGN_v2}
\includegraphics[width=0.5\textwidth,angle=0]{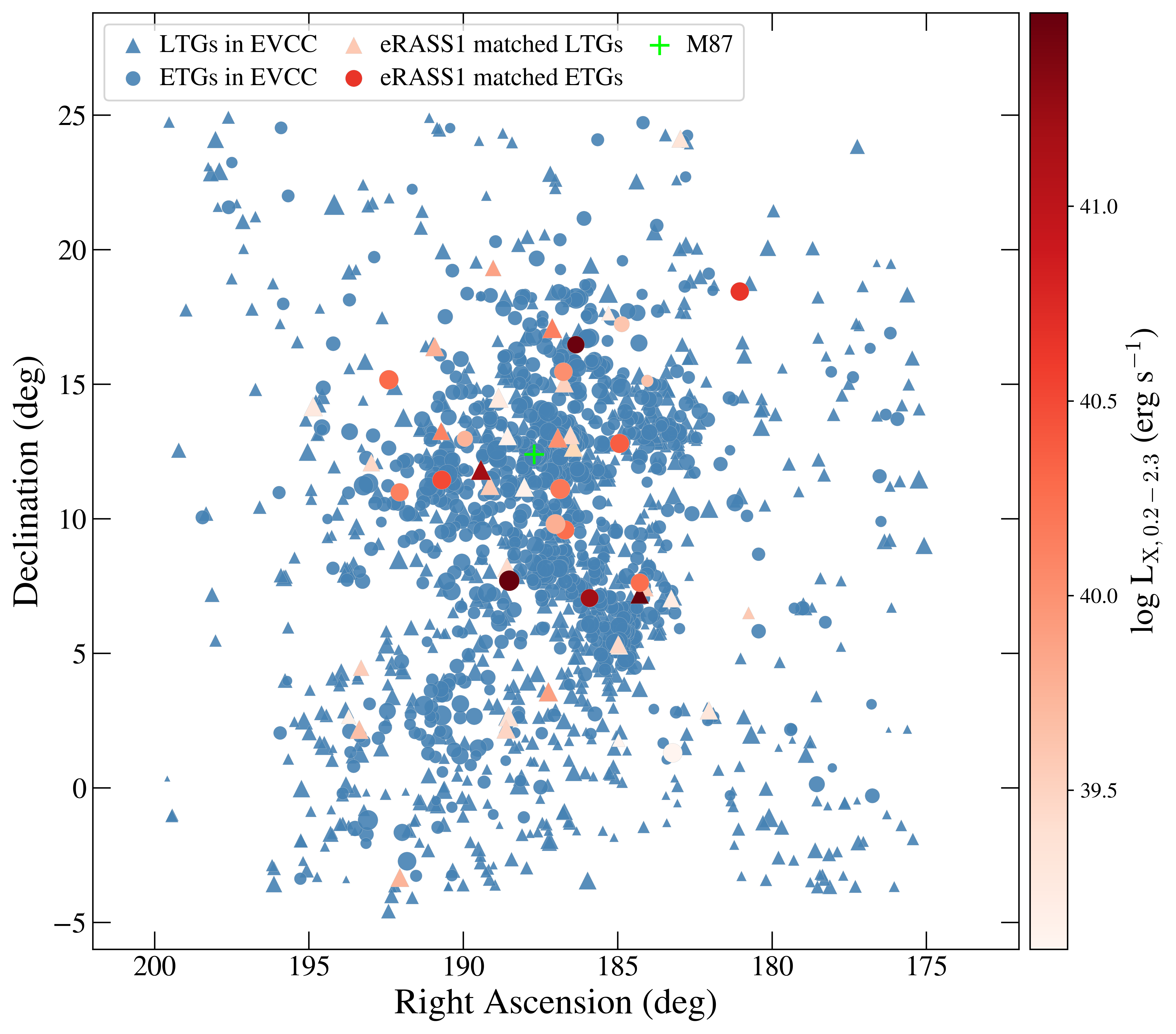}
\includegraphics[width=0.5\textwidth,angle=0]{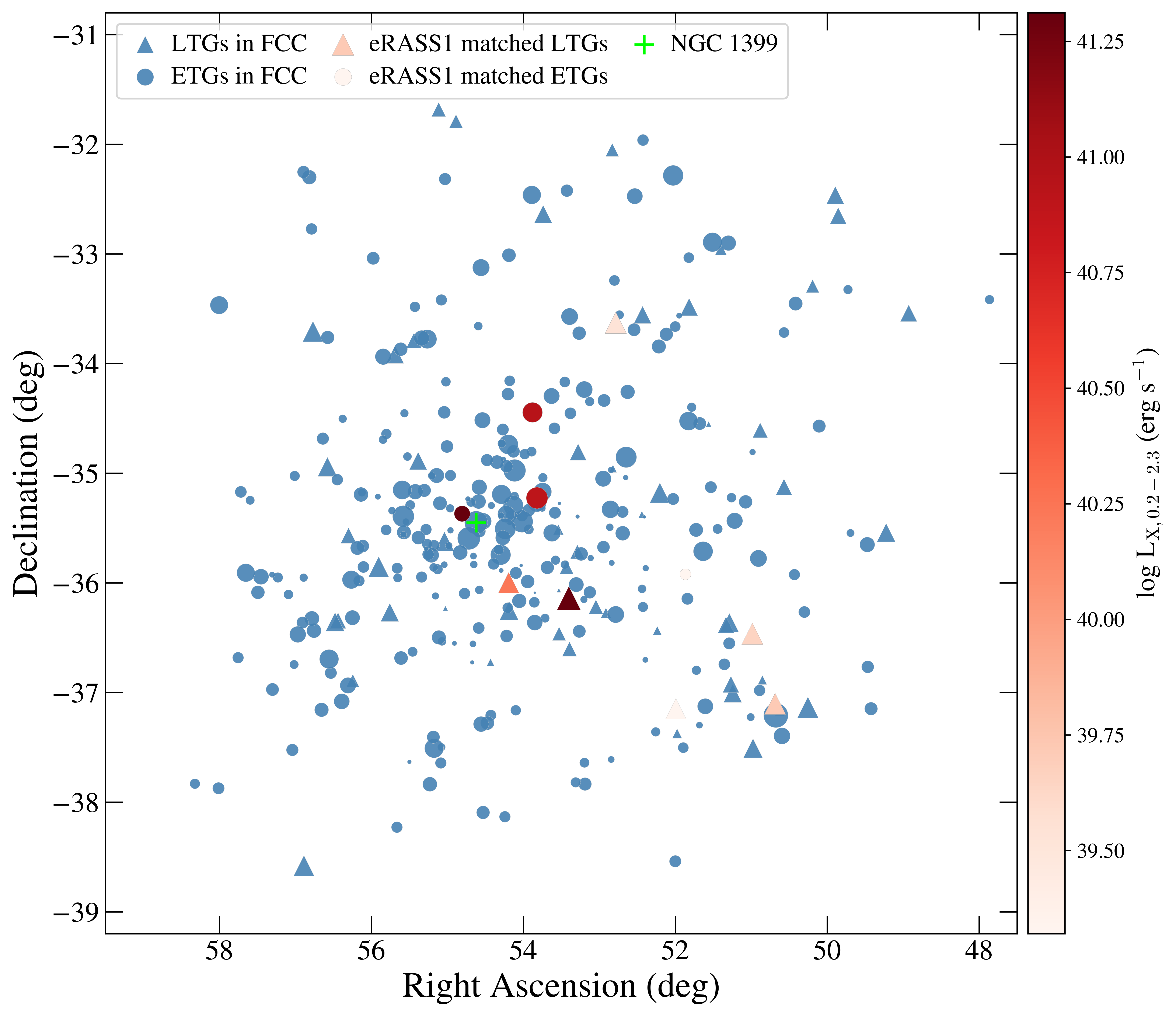}
\caption{%eROSITA image of the Virgo and Fornax cluster. 
The sky distribution of member galaxies in Virgo ({\it left}) and Fornax ({\it right}). 
Member galaxies in the Extended Virgo Cluster Catalog and Fornax Cluster Catalog are indicated by blue symbols. The nuclear X-ray sources detected by eRASS1 are marked by red symbols color-coded by the 0.2--2.3 keV luminosity. ETGs and LTGs are represented by circles and triangles. The size of each symbol is scaled with stellar mass (approximated by $B_T$ mag for the Fornax galaxies).
\label{fig:Virgo}}
\end{figure*}

\section{Member Galaxies of Virgo and Fornax}\label{sec:catalog}
To identify X-ray AGNs in the two clusters, we employ the Extended Virgo Cluster Catalog (EVCC; \citealp{2014ApJS..215...22K}) and the Fornax Cluster Catalog (FCC; \citealp{1989AJ.....98..367F}), which provide the most up-to-date, uniformly classified member galaxies of the respective clusters. 

The EVCC contains a total of 1589 spectroscopically identified galaxies, based primarily on the Sloan Digital Sky Survey (SDSS) and supplemented by spectroscopic data from the literature. 
The footprint of EVCC is more than 5 times that of the photography-based Virgo Cluster Catalog \citep{1985AJ.....90.1681B} and reaches out to a projected distance 3.5 times the virial radius of Virgo ($R_{200} \approx 1.0$ Mpc;  \citealp{2017MNRAS.469.1476S}).
As detailed in \citet{2014ApJS..215...22K}, within this footprint, nearly 100\% of the SDSS photometric galaxies with a $r$-band magnitude $r \lesssim 14$ mag have a spectroscopic redshift, and the 50\% completeness level for galaxies having a radial velocity $< 3000\rm~km~s^{-1}$ (i.e., the criterion for a Virgo member) is at $r \sim 16.5$ mag. The latter value translates to an absolute magnitude of $M_r \approx -14.1$ for the nominal distance of Virgo (16.5 Mpc; \citealp{2007ApJ...655..144M}), 
%distance modulus 31.09
roughly corresponding to a stellar mass of $\lesssim 10^8\rm~M_\odot$. 
Generally speaking, few (S)MBHs are expected to exist in galaxies with a similar stellar mass or lower. Therefore, the EVCC should contain the majority, if not all, of member galaxies within which an X-ray AGN could be present. 

The FCC, on the other hand, is a photography-based catalog containing 340 likely members in a footprint of $6^\circ \times 6^\circ$ (2.0 Mpc $\times$ 2.0 Mpc; at the nominal distance of 20.0 Mpc of Fornax; \citealp{2009ApJ...694..556B}), reaching out to $\sim1.5$ times the virial radius ($R_{200} \approx 0.7$ Mpc; \citealp{2001MNRAS.326.1076D}).
The FCC has a completeness limit of $B_{\rm T} \sim 18$ mag (an absolute magnitude of ${M_{B_{\rm T}}} \sim -13.5$), which corresponds to a stellar mass $\sim 10^8\rm~M_\odot$. This means that the FCC should also include all member galaxies likely to host an AGN, with the caveat that the membership of most galaxies was not on a spectroscopic basis. 
% distance modulus 31.51

%Stellar mass and morphological type
%DESI or SDSS images

\begin{figure*}[ht!]
\epsscale{1}
\includegraphics[width=0.5\textwidth]{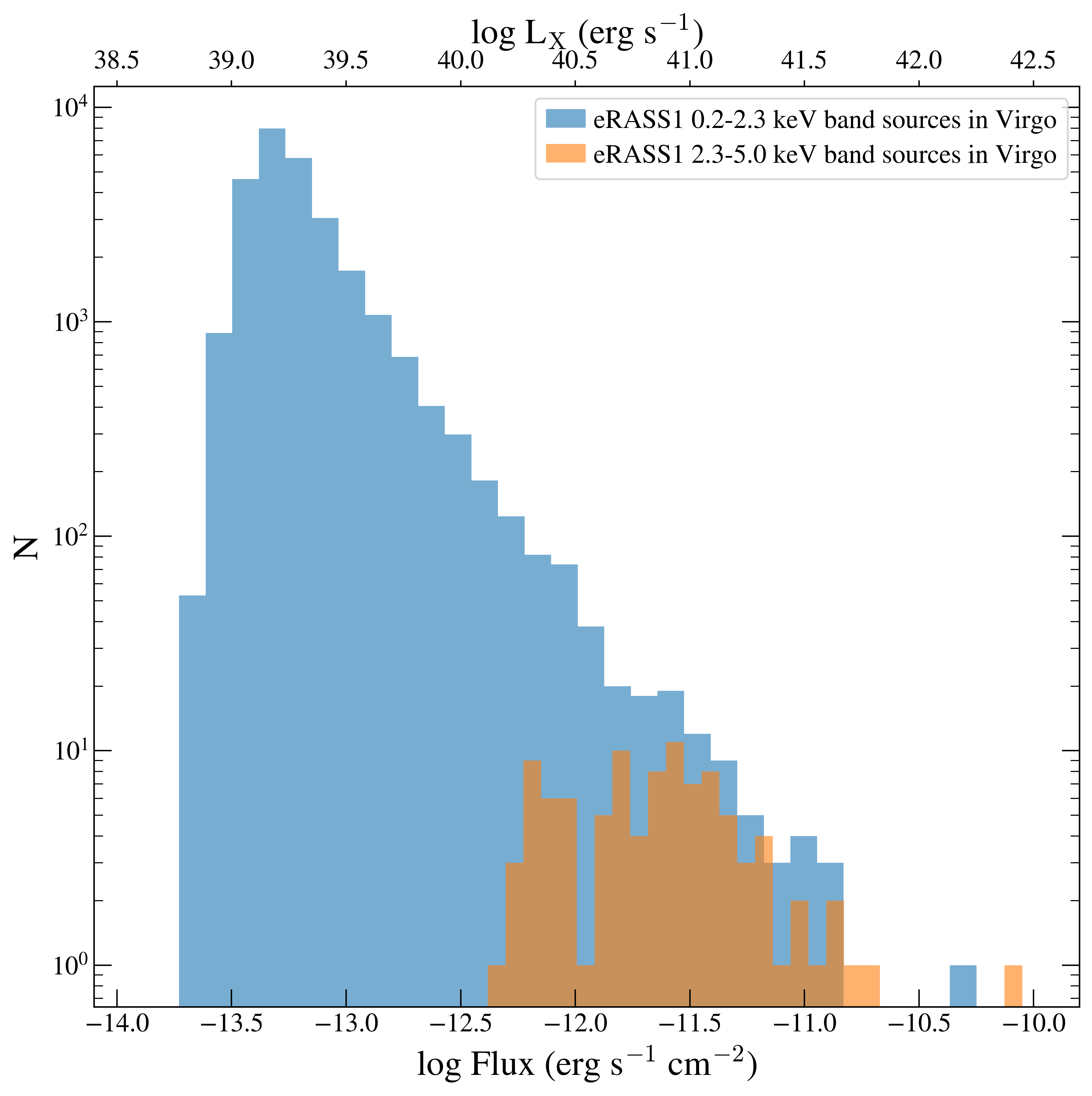}
\includegraphics[width=0.5\textwidth]{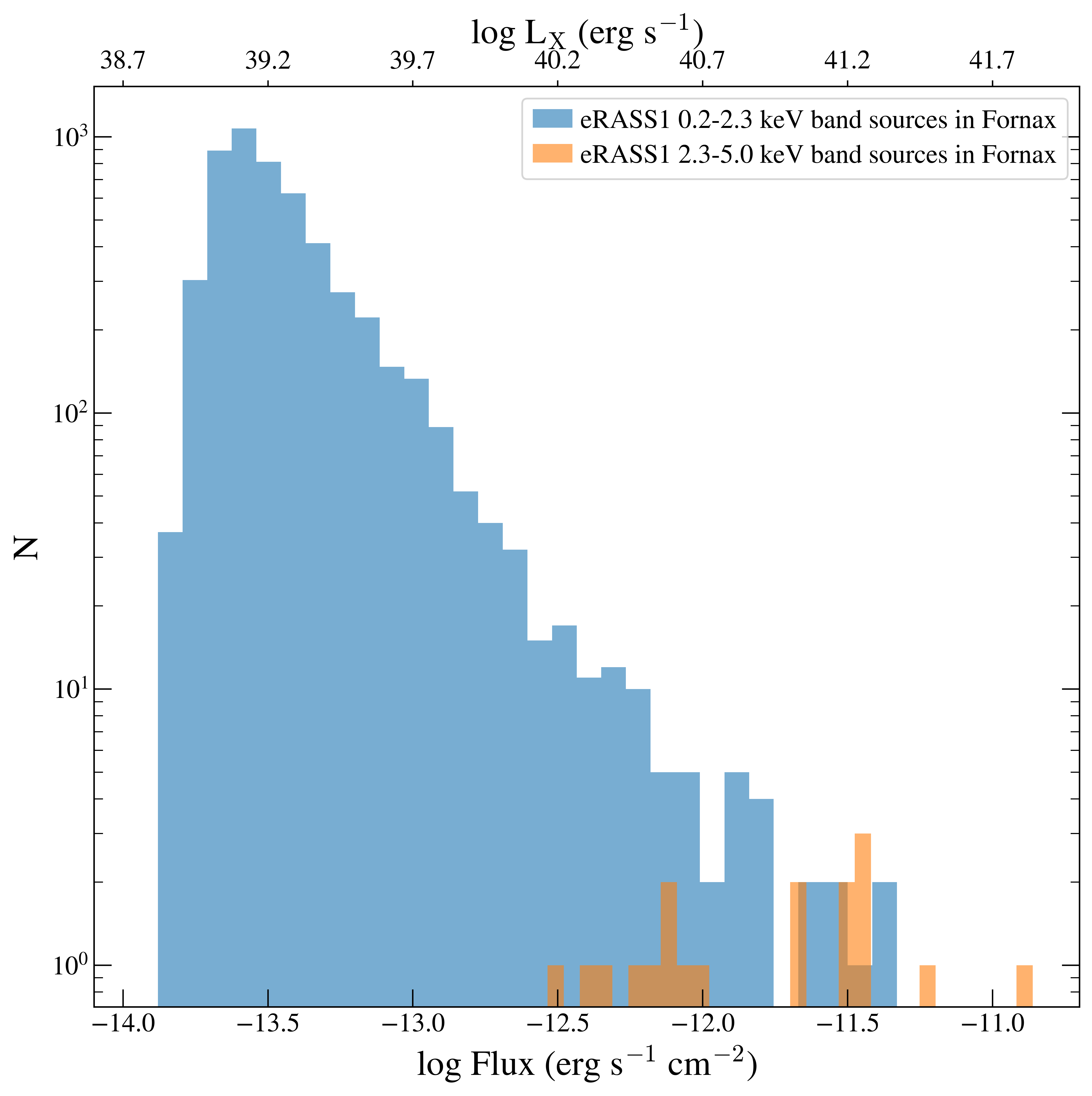}
\caption{X-ray flux and luminosity distributions of all the eRASS1 point sources in the Virgo ({\it left})and Fornax ({\it right}) footprint. Soft band and hard band sources are shown in blue and orange histograms. The peak of the distributions indicates the detection completeness limit.
\label{fig:sensitivity}}
\end{figure*}

\section{Identification of nuclear X-ray sources}\label{sec:nucleus}

To identify a putative X-ray AGN, we cross-match the galactic nuclei, whose sky coordinates are provided by the EVCC and FCC, with the newly released all-sky survey X-ray source catalogs from eRASS1 \citep{2024A&A...682A..34M}, which include a soft-band (0.2--2.3 keV) catalog and a hard-band (2.3--5 keV) catalog.
We adopt a matching radius of $10\arcsec$, corresponding to the 99 percentile of the position error of the soft-band sources,
which spans a linear scale of $\sim$0.8 ($\sim$1.0 kpc) at the distance of Virgo (Fornax).

For Virgo, we find a total of 50 point-like X-ray sources positionally coincident with the optical nuclei (Figure~\ref{fig:Virgo}).
Among them, 49 are from the soft band, with 0.2--2.3 keV luminosities ($L_{\rm 0.2-2.3}$) ranging between $9.2\times10^{38}\rm~erg~s^{-1} - 3.1\times10^{41}\rm~erg~s^{-1}$. 
We take the peak of the 0.2--2.3 keV flux distribution of all eRASS1 sources detected within the EVCC footprint as a rough indicator of the detection completeness limit, which corresponds to $\sim 1.6\times10^{39}\rm~erg~s^{-1}$ at the distance of Virgo (Figure~\ref{fig:sensitivity}).
This value is also roughly consistent with the aperture photometry based sensitivity map provided by eRASS1, which suggests a 50\% completeness limit of $\sim 6.6\times10^{-14}\rm~erg~s^{-1}~{cm^{-2}}$ ($\sim 2.2\times10^{39}\rm~erg~s^{-1}$).
Three sources (EVCC\,270, 576, 965) are detected in the hard band, among which one source (EVCC\,576 = NGC\,4388) is detected in the hard band only. The 2.3--5 keV luminosity of these sources is $6.0\times10^{41}, 9.2\times10^{40}$ and $2.3\times10^{41}\rm~erg~s^{-1}$, respectively.

We estimate the number of potential random matches by artificially shifting the centroid of all X-ray sources within the EVCC footprint by $\pm3\arcmin$ in either $R.A.$ or $Decl.$, an amount sufficiently large compared to the default matching radius but also sufficiently small compared to the angular range of the member galaxies. This exercise results in a random match of 1.0 sources averaged over the four directions, which may account for only 2\% of the actual matches. 

In the case of Fornax, originally 11 nuclear X-ray sources were found. After consulting with the NASA/IPAC Extragalactic Database (NED)\footnote{https://ned.ipac.caltech.edu}, we find that one galaxy (FCC\,129) has a large redshift that disputes its Fornax membership, while the other 10 galaxies have a redshift compatible with Fornax. 
Therefore, we retain a total of 10 nuclear sources (Figure~\ref{fig:Virgo}), 
which are all detected in the soft band, with $L_{\rm 0.2-2.3}$ ranging between $9.6\times10^{38}\rm~erg~s^{-1} - 2.1\times10^{41}\rm~erg~s^{-1}$. 
We note that the eRASS1 source detection completeness limit is $\sim 1.2\times10^{39}\rm~erg~s^{-1}$ within the footprint of FCC (Figure~\ref{fig:sensitivity}). %This is somewhat lower than that of Virgo, presumably due to an on-average longer survey exposure for Fornax's higher ecliptic latitude. 
Only one galaxy (FCC\,121 = NGC\,1365) is detected in the hard band with a 2.3--5 keV luminosity of $6.6\times10^{41}\rm~erg~s^{-1}$.
We find 0.5 random matches for the Fornax galaxies. 
%Figure~\ref{fig:Virgo} displays the positions of the 50 matched nuclear X-ray sources in Virgo, against the positions of all EVCC galaxies. 
%which may account for only 5\% of the actual matches.
%We find that random matches are also negligible due to the Fornax galaxies having a similar surface density as their Virgo counterparts.

Table~\ref{table:AGN} lists the host galaxy name, centroid coordinates, X-ray luminosity of the matched eRASS1 source, and stellar mass. The stellar mass of the Virgo galaxies is calculated based on the SDSS $g$- and $r$-band photometry given by the EVCC and the color-magnitude relation of \citet{2003ApJS..149..289B}.
Since only a single band photometry is provided by the FCC, we instead employ the WISE \citep{2010AJ....140.1868W} W1 and W2 band images to calculate the stellar mass, following the calibration of \citet{2019ApJS..245...25J}.
We have adopted the nominal distances of Virgo and Fornax when deriving the X-ray luminosity and stellar mass.

{\it Comparison with \citet{2010ApJ...714...25G}.} The AMUSE-Virgo program detected 32 nuclear X-ray sources out of 100 ETGs in Virgo down to a limiting luminosity of $3.7\times10^{38}\rm~erg~s^{-1}$ over 0.5--7 keV (corresponds to $3.2\times10^{38}\rm~erg~s^{-1}$ in the 0.2--2.3 keV by assuming an absorbed power-law model with the photon index of 2 and column density  $N_{\rm H}$ = $2.5 \times 10^{20} {\rm~cm^{-2}}$ adopted in \citealp{2010ApJ...714...25G}). 
We find 3 eRASS1 detections among these 32 targets (EVCC\,353, 681, 2211), while among the 68 non-detected ETGs, two (EVCC\,626, 884) have an eRASS1 match.
The non-detection of the remaining targets can be mainly attributed to a lack of sensitivity, because most of them had an X-ray luminosity below $10^{39}\rm~erg~s^{-1}$ according to \citet{2010ApJ...714...25G}. Particularly worth mentioning is M\,87, the nuclear source of which has an X-ray luminosity $\sim 10^{41}\rm~erg~s^{-1}$ but is still undetected by eRASS1. This is likely due to the heavy contamination by the surrounding diffuse hot gas in this brightest cluster galaxy.

{\it Comparison with \citet{2022MNRAS.512.3284S}.} \citet{2022MNRAS.512.3284S} reported that a total of 39 nuclear X-ray sources are present among the 75 LTGs in Virgo with a typical detection limit of $\sim 3\times10^{38}\rm~erg~s^{-1}$ and a completeness limit of $\sim 10^{39}\rm~erg~s^{-1}$ in the 0.3--10 keV band, but without quoting the exact host galaxies. As a byproduct of our recent study of diffuse hot gas in these LTGs \citep{2024ApJ...961..249H}, we have independently detected 35 nuclear X-ray sources. Among them, 13 share the same host galaxy as the eRASS1 nuclear sources. 
One additional galaxy, EVCC\,595, has an eRASS1 detection.
%The non-detection of the remaining targets is presumably due to a lack of sensitivity with eRASS1.

{\it Comparison with \citet{2019ApJ...874...77L}.} \citet{2019ApJ...874...77L} detected a total of 11 out of 29 ETGs in Fornax down to a limiting luminosity of $5\times10^{38}\rm~erg~s^{-1}$ over 0.3--10 keV, all having an X-ray luminosity below $\sim 2\times10^{39}\rm~erg~s^{-1}$. We find only one eRASS1 match (with FCC\,153) among these 11 targets and one additional eRASS1 match (with FCC\,147) among the 18 non-detected galaxies.
%The non-detection of the remaining targets can also be understood as a lack of sensitivity with eRASS1.

%black hole mass?

\startlongtable
\begin{deluxetable}{ccccccc}
\tablenum{1}
\tablecaption{Nuclear X-ray sources and host galaxy properties}
\tablewidth{0pt}
\tabletypesize{\scriptsize}
\tablehead{
\colhead{Name} & \colhead{ NGC }& RA & DEC & \colhead{Morph. } &\colhead{$\log L_\mathrm{X}$}  & \colhead{$\log M_\star$} \\
& &   ($\mathrm{deg}$) &   ($\mathrm{deg}$) &  & ($\mathrm{erg\ s^{-1}}$) & ($\mathrm{M_\odot}$) 
}
\decimalcolnumbers
\startdata
EVCC & (Virgo) \\
\hline
86   & --   & 180.7616 &  6.5014 & Irr & $39.58^{+0.13}_{-0.14}$ &  7.9 \\ 
93   & 4064 & 181.0469 & 18.4442 & SB0      & $39.47^{+0.15}_{-0.18}$ & 10.2 \\ 
107  & 4123 & 182.0464 &  2.8786 & SBc      & $39.30^{+0.19}_{-0.22}$ & 10.1 \\ 
153  & 4162 & 182.9690 & 24.1233 & Sc       & $39.33^{+0.19}_{-0.22}$ &  9.8 \\ 
176  & 4180 & 183.2627 &  7.0389 & Sb       & $39.23^{+0.20}_{-0.25}$ &  9.9 \\ 
241  & --   & 184.0418 & 15.1236 & dE       & $39.14^{+0.20}_{-0.26}$ &  7.8 \\ 
251  & 4224 & 184.1408 &  7.4619 & Sa       & $39.60^{+0.20}_{-0.09}$ & 10.3 \\ 
267  & 4233 & 184.2820 &  7.6243 & SB0      & $39.34^{+0.18}_{-0.19}$ & 10.2 \\ 
270  & 4235 & 184.2912 &  7.1915 & Sa       & $41.50^{+0.01}_{-0.02}$ & 10.3 \\ 
349  & --   & 184.8695 & 17.2305 & dS0   & $39.13^{+0.21}_{-0.25}$ &  9.0 \\ 
353  & 4267 & 184.9386 & 12.7982 & SB0      & $39.38^{+0.16}_{-0.18}$ & 10.6 \\ 
355  & 4268 & 184.9467 &  5.2837 & Sa       & $39.33^{+0.14}_{-0.16}$ &  9.8 \\ 
357  & --   & 184.9713 &  1.7733 & Irr & $39.09^{+0.21}_{-0.25}$ &  8.2 \\ 
359  & 4273 & 184.9833 &  5.3424 & SBc      & $39.45^{+0.12}_{-0.13}$ &  9.8 \\ 
399  & --   & 185.2976 & 17.6387 & Irr & $39.22^{+0.19}_{-0.21}$ &  8.6 \\ 
489  & 4342 & 185.9125 &  7.0540 & S0       & $39.65^{+0.10}_{-0.11}$ & 10.1 \\ 
555  & 4383 & 186.3561 & 16.4699 & S0       & $39.73^{+0.12}_{-0.12}$ &  9.8 \\ 
576* & 4388 & 186.4442 & 12.6629 & Sa       & $\lesssim$ 39.7 & 10.4 \\ 
595  & 4402 & 186.5306 & 13.1125 & Sb       & $39.46^{+0.16}_{-0.18}$ & 10.2 \\ 
626  & 4417 & 186.7103 &  9.5844 & S0       & $39.34^{+0.13}_{-0.15}$ & 10.5 \\ 
634  & 4419 & 186.7348 & 15.0476 & Sa       & $39.52^{+0.14}_{-0.16}$ & 10.5 \\ 
637  & 4421 & 186.7606 & 15.4616 & SB0      & $39.26^{+0.19}_{-0.20}$ & 10.2 \\ 
673  & 4438 & 186.9381 & 13.0110 & Sb       & $40.01^{+0.09}_{-0.09}$ & 10.8 \\ 
681  & 4442 & 187.0162 &  9.8037 & SB0      & $39.19^{+0.16}_{-0.19}$ & 10.8 \\ 
884  & 4526 & 188.5127 &  7.6993 & S0       & $39.74^{+0.10}_{-0.11}$ & 11.2 \\ 
886  & 4527 & 188.5357 &  2.6529 & Sb       & $39.37^{+0.16}_{-0.19}$ & 10.8 \\ 
889  & 4531 & 188.5662 & 13.0754 & Sa       & $39.15^{+0.17}_{-0.20}$ & 10.2 \\ 
892  & 4535 & 188.5844 &  8.1982 & SBc      & $39.44^{+0.15}_{-0.17}$ & 10.6 \\ 
930  & 4561 & 189.0339 & 19.3227 & SBd      & $39.88^{+0.09}_{-0.10}$ &  9.3 \\ 
938  & 4568 & 189.1431 & 11.2394 & Sc       & $39.58^{+0.13}_{-0.14}$ & 10.5 \\ 
965  & 4579 & 189.4310 & 11.8182 & SBb      & $41.20^{+0.02}_{-0.02}$ & 11.1 \\ 
1006 & --   & 189.9501 & 12.9739 & S0       & $39.18^{+0.18}_{-0.22}$ &  9.2 \\ 
1080 & 4639 & 190.7185 & 13.2571 & SBb      & $40.11^{+0.07}_{-0.07}$ &  9.9 \\ 
1102 & 4651 & 190.9280 & 16.3938 & Sc       & $39.77^{+0.10}_{-0.10}$ & 10.4 \\ 
1170 & 4691 & 192.0565 & -3.3323 & SBa      & $39.75^{+0.09}_{-0.11}$ & 10.3 \\ 
1171 & 4694 & 192.0628 & 10.9835 & S0       & $39.30^{+0.30}_{-0.11}$ & 10.1 \\ 
1182 & 4710 & 192.4126 & 15.1657 & S0       & $39.35^{+0.17}_{-0.18}$ & 10.6 \\ 
1202 & 4746 & 192.9801 & 12.0830 & Sc       & $39.46^{+0.14}_{-0.16}$ &  9.7 \\ 
1216 & 4765 & 193.3105 &  4.4634 & Sm       & $39.56^{+0.12}_{-0.14}$ &  9.2 \\ 
1237 & 4809 & 193.7128 &  2.6525 & Irr & $39.18^{+0.18}_{-0.23}$ &  8.7 \\ 
2034 & 4179 & 183.2171 &  1.2997 & S0       & $38.96^{+0.21}_{-0.24}$ & 10.5 \\ 
2111 & 4429 & 186.8607 & 11.1077 & S0       & $39.32^{+0.15}_{-0.18}$ & 10.9 \\ 
2124 & 4450 & 187.1218 & 17.0846 & Sb       & $40.13^{+0.06}_{-0.07}$ & 10.9 \\ 
2129 & 4457 & 187.2459 &  3.5706 & Sa       & $39.90^{+0.08}_{-0.09}$ & 10.6 \\ 
2154 & 4503 & 188.0260 & 11.1764 & Sa       & $39.15^{+0.20}_{-0.23}$ & 10.5 \\ 
2169 & 4536 & 188.6130 &  2.1878 & Sc       & $39.48^{+0.13}_{-0.16}$ & 10.5 \\ 
2174 & 4548 & 188.8610 & 14.4955 & SBb      & $39.25^{+0.14}_{-0.16}$ & 10.8 \\ 
2211 & 4638 & 190.6976 & 11.4425 & S0       & $39.42^{+0.15}_{-0.18}$ & 10.4 \\ 
2240 & 4772 & 193.3713 &  2.1684 & Sa       & $39.66^{+0.11}_{-0.11}$ & 10.4 \\ 
2254 & 4866 & 194.8628 & 14.1712 & Sa       & $39.26^{+0.17}_{-0.19}$ & 10.5 \\
\hline
FCC  & (Fornax) \\
\hline
22  & 1317 & 50.68521 & -37.10322 & Sa   & $39.72^{+0.09}_{-0.10}$ & 10.6  \\ 
29  & 1326 & 50.98410 & -36.46461 & SBa   & $39.65^{+0.09}_{-0.10}$ & 10.7 \\ 
57  & --   & 51.86572 & -35.92211 & dE6      & $38.98^{+0.19}_{-0.23}$ & 7.7 \\ 
62  & 1341 & 51.99234 & -37.14789 & Sbc   & $39.32^{+0.13}_{-0.15}$ & 9.6 \\ 
88  & 1350 & 52.78314 & -33.62839 & SBb & $39.53^{+0.10}_{-0.12}$ & 10.9 \\ 
121 & 1365 & 53.40071 & -36.13970 & SBbc & $41.31^{+0.01}_{-0.01}$ & 10.6 \\ 
147 & 1374 & 53.81984 & -35.22605 & E0       & $39.22^{+0.16}_{-0.20}$ & 10.6 \\ 
153 & --   & 53.87885 & -34.44572 & S0       & $39.22^{+0.15}_{-0.17}$ & 10.2 \\ 
179 & 1386 & 54.19280 & -35.99919 & Sa       & $40.23^{+0.04}_{-0.05}$ &  9.6 \\ 
222 & --   & 54.80533 & -35.36965 & dE0    & $39.28^{+0.16}_{-0.20}$ &  8.6
\enddata
%\resizebox{\textwidth}{!}{
%\tablecomments{(1) Galaxy Name as in the EVCC and FCC catalogs. (2) NGC name of the galaxy. (3)--(4) Right ascension and declination at equinox J2000 of the galactic nuclei. (5) The morphological type of the galaxy. (6) X-ray luminosity and 1$\sigma$ error in the 0.2--2.3 keV energy band. $^*$Non-detected. (7) Stellar mass of the galaxy. For Virgo galaxies, this is derived from the $g-r$ mass-luminosity relation, with a typical uncertainty of $\sim 0.1$ dex. For Fornax galaxies, this is derived from WISE W1- and W2-band images, with a typical uncertainty of $\sim0.1-0.2$ dex. FCC\,57 is too faint to be reliably measured in the WISE images, hence its stellar mass a rough estimate from its $B_{\rm T}$ mag.}
%}
\label{table:AGN}
\end{deluxetable}
{\vskip-1cm}
\noindent{Note--\footnotesize(1) Galaxy Name as in the EVCC and FCC catalogs. (2) NGC name of the galaxy. (3)--(4) Right ascension and declination at equinox J2000 of the galactic nuclei. (5) The morphological type of the galaxy. (6) X-ray luminosity and 1$\sigma$ error in the 0.2--2.3 keV energy band. $^*$Non-detected. (7) Stellar mass of the galaxy. For Virgo galaxies, this is derived from the $g-r$ mass-luminosity relation, with a typical uncertainty of $\sim 0.1$ dex. For Fornax galaxies, this is derived from WISE W1- and W2-band images, with a typical uncertainty of $\sim0.1-0.2$ dex. FCC\,222 has a larger uncertainty of $1$ dex due to its low luminosity. FCC\,57 is too faint to be reliably measured in the WISE images, hence its stellar mass a rough estimate from its $B_{\rm T}$ mag.}

\section{Statistical properties}\label{sec:stat}
\subsection{Potential contamination by host galaxy}
\label{subsec:contam}

Due to the moderate resolution of eROSITA, the detected nuclear X-ray sources might be contaminated by non-nuclear X-ray emission of the host galaxy. 
In particular, low mass X-ray binaries (LMXBs), the amount and total X-ray luminosity of which scale with the stellar mass, is expected to produce a 0.2--2.3 keV luminosity of
$L_{\rm LMXB} \approx 7.9\times10^{38}\rm~erg~s^{-1}(M_*/10^{10}\rm~M_\odot$),
where $M_*$ is the stellar mass enclosed within the eRASS1 photometry aperture ($\sim 30\arcsec$). 
The above scaling relation is derived from the empirical $L$(2--10 keV)--$M_*$ relation of \citet{2010ApJ...724..559L}, and we have assumed for LMXBs an intrinsic power-law spectrum with a canonical photon-index of 1.7. 
Based on the EVCC, we find that
the majority of matched galaxies have an optical size ($r_{\rm Kron}$) above $30\arcsec$ (i.e., well resolved by eROSITA); only a few galaxies having the lowest stellar masses ($\lesssim 10^9\rm~M_\odot$) are entirely covered by the eRASS1 source aperture. 
Therefore, we conclude that statistically LMXBs have only a small contribution to the detected nuclear X-ray emission.

A small fraction of the matched galaxies could be actively forming stars, which would also contribute to the observed X-ray emission in the form of high-mass X-ray binaries and/or diffuse hot gas. The 0.2--2.3 keV luminosity of this star formation-related component is estimated as \citep{2016ApJ...825....7L}: $L_{\rm SF} \approx 5.6\times10^{39}\rm~erg~s^{-1}(SFR/\rm M_\odot~yr^{-1}$), for which a factor of 1.4 is multiplied to the original scaling relation to roughly account for a different photon energy range (0.5--2 keV). Since no explicit information about the star formation rate (SFR) is provided by the EVCC, we employ the WISE W3-band image to estimate the SFR, following \citet{2019ApJS..245...25J}.
We find that all but two of the 50 host galaxies in Virgo have an SFR within the eRASS1 source aperture too low to have a significant contribution to the observed nuclear X-ray emission. 
The two exceptions are EVCC\,359 and EVCC\,892 which have a central SFR $\sim 1.7\rm~M_\odot~yr^{-1}$ and $\sim 0.4\rm~M_\odot~yr^{-1}$, sufficient to explain the observed nuclear X-ray emission. 
Therefore, we conclude that statistically star-forming activities also have only a minor contribution to the detected nuclear X-ray emission.

By evaluating the stellar mass and SFR of the 10 Fornax host galaxies, we find that host galaxy contamination can be neglected in all but three galaxies. FCC\,22, FCC\,29 and FCC\,62 have a substantial SFR to account for the observed X-ray emission. 
%a similar conclusion can be drawn about the host galaxy contamination.
Moreover, it is still possible that a small fraction of the eRASS1 nuclear sources are X-ray binaries rather than a genuine AGN, because the detection limit of eRASS1 is still compatible with the Eddington limit of stellar-mass black holes. We note that the same caveat holds for the aforementioned {\it Chandra} programs, which all have an even lower detection limit.

\subsection{Host galaxies and occupation fraction}
\label{subsec:host}
%The morphological type of the 50 host galaxies in Virgo is taken from the EVCC, except for EVCC\,595 (= NGC\,4402) and EVCC\,1182 (= NGC\,4710) where the original classification of `edge-on' is replaced by the morphological type from NED. 
%Similarly, the morphological type of the 10 Fornax host galaxies is taken from the FCC.

Following the convention, we collectively assign ellipticals, lenticulars and dwarf ellipticals/lenticulars as ETGs, and the remaining morphological types as LTGs.
For Virgo, when the EVCC morphological type is `edge-on', we consult with NED to determine the morphological type if available; for those without an explicit morphological type, we take them as ETGs, introducing only a $\sim2.5\%$ ambiguity. 
The ETG fraction of Virgo is thus $\sim$49\%. A much higher fraction of $\sim$82\% is found in Fornax according to the FCC, with the caveat that the FCC classification was based on old imaging data probably less sensitive than SDSS. 
It turns out that 17 of the 50 Virgo hosts are ETGs, while 
4 of the 10 Fornax hosts are ETGs, indicating a similar percentage ($34\%\pm10\%$ vs. $40\%\pm24\%$) in the two clusters. 
Quoted errors are of Poisson and at $1\sigma$ confidence level.

%These values are to be contrasted with the fraction of ETGs in the two clusters. 

The percentage of X-ray detected nuclei, or the {\it occupation rate} of an X-ray AGN, is also quite similar between Virgo and Fornax (50/1589 $\approx~3.1\%\pm0.4\%$ vs. 10/340 $\approx~2.9\%\pm0.9\%$).
The occupation rate becomes $\approx~2.6\%\pm0.4\%$ and $\approx~2.6\%\pm0.9\%$, if we drop 9 sources in Virgo and 1 source in Fornax, whose X-ray luminosities are below the detection completeness limit (Section~\ref{sec:nucleus}).
If instead we restrict on host galaxies with ${\rm log}(M_*/{\rm M_\odot}) > 7.8$, which is the lowest mass among the 50 Virgo hosts and also is roughly the completeness limit of the EVCC, the percentage becomes $4.2\pm0.6\%$ (50/1203).

We further evaluate the occupation rate by dividing the Virgo galaxies into subgroups, according to the stellar mass, the projected distance from the center of M\,87, and the morphological type (ETG or LTG), respectively.
%The resultant occupation rates are summarized in Table~\ref{table:occupation}.
Due to the small number of X-ray nuclei in Fornax, we do not perform a similar analysis.

As expected, while the majority of EVCC galaxies are actually dwarfs, only four X-ray nuclei are detected among galaxies with $7.8 < {\rm log}(M_*/{\rm M_\odot}) < 8.65$ (4/602 $\approx 0.7\pm0.3\%$), compared to 46 in the more massive group (46/601 $\approx 7.7\pm1.1\%$).
On the other hand, a comparable occupation rate ($3.6\pm0.7\%$ vs. $2.6\pm0.6\%$) is found between galaxies located inside and outside $\sim 1.7 R_{200}$ of Virgo. Intuitively, those galaxies on the first infall to the cluster core may still contain a significant amount of ISM to fuel the central SMBH, but this effect could be offset by the on-average heavier SMBHs in galaxies already in or passing the cluster core due to frequent galaxy merger therein. A full spectroscopic survey of the nuclei of the EVCC galaxies would be desired to provide a robust and uniform measurement of black hole mass and hence the Eddington ratio. 
Finally, a marginally significant ($\sim2\sigma$) difference in the occupation rate is seen between the ETGs and LTGs ($2.2\pm0.5\%$ vs. $4.0\pm0.7\%$). 
This is consistent with the naive expectation that LTGs contain more available fuel for the SMBH.

%{\bf [some description about specific cases, e.g. dwarf hosts]}

%\subsection{Nuclear X-ray Luminosity Function}\label{sec:XLF}
%Figure~2 displays the X-ray Luminosity Functions (XLF) of the matched nuclei in both Virgo and Fornax. 
%For comparison, we also plot the XLFs of the AMUSE-Virgo ETGs, Virgo LTGs, Fornax ETGs, Antlia ETGs, as well as field ETGs, which are all based on {\it Chandra} observations. The detection limits of these surveys are different. 

%\begin{deluxetable}{cccc}
%\tablenum{2}
%\tablecaption{Occupation rate of Virgo X-ray AGN}
%\tablewidth{0pt}
%\tabletypesize{\scriptsize}
%\tablehead{
%\colhead{Group} &  \colhead{Stellar mass} &\colhead{Projected radius} &  \colhead{Morphology}
%}
%%\decimalcolnumbers
%\startdata
%$7.8\leq $log$M_* \leq 8.65$ & 4/602 (0.7\pm0.3\%) & - &  -  \\
%log$M_* > 8.65 $ & 46/601 (7.7\pm1.1\%) & - & - \\
%\hline
%$R < 1.68$ Mpc & - &  28/795 (3.6\pm0.7\%)   &  -  \\
%$R > 1.68$ Mpc & - &  22/794 (2.6\pm0.6\%)  & - \\
%\hline
%ETG & -& - &  17/774 (2.2\pm0.5\%)  \\
%LTG & - & - & 33/815 (4.0\pm0.7\%) \\
%\hline
%\enddata
%\tablecomments{Quoted errors are at 68\% confidence level.}
%\label{table:occupation}
%\end{deluxetable}
%Data Generated From AN30S2_v7

\section{Summary and discussion} \label{sec:sum}
In this work, we have utilized the newly released X-ray source catalogs from the first all-sky scan of eROSITA to probe putative X-ray AGNs in the Virgo and Fornax clusters, taking advantage of its full coverage of all known member galaxies. 

A total of 50 and 10 X-ray sources are found to be positionally coincident with the nuclei of member galaxies in Virgo and Fornax, respectively. A large fraction of these nuclear X-ray sources are identified for the first time. 
Despite the moderate resolution of eROSITA, we find that in most cases host galaxy contamination is negligible to the observed X-ray luminosity, indicating the dominance of a genuine X-ray AGN.
The majority of these nuclear X-ray sources are  
found in late-type hosts.

The overall occupation rate of the eRASS1 nuclear sources is remarkably low, $\sim3\%$ in both clusters. This appears to be much lower than found in nearby galaxies not residing in clusters. For instance, at least 13\% of the AMUSE-Field ETGs have an X-ray luminosity $\gtrsim 2\times10^{39}\rm~erg~s^{-1}$ \citep{2012ApJ...747...57M}, i.e., above the eRASS1 completeness limit. 
In a {\it Chandra} survey of X-ray AGNs in 719 galaxies within a distance of 50 Mpc (most are outside Virgo and Fornax), \citet{2017ApJ...835..223S} found 314 X-ray nuclei, among which $\sim50\%$ have an X-ray luminosity above the eRASS1 completeness limit.
Although not free of selection effect, these two studies imply that the occupation rate of an X-ray AGN in field galaxies is likely higher than 10\%. 
Hence, the relatively low occupation rate in Virgo and Fornax strongly suggests that AGN activity is generally suppressed in the cluster environment. 

A related and equally remarkable feature about the eRASS1 nuclear X-ray sources is the paucity of luminous objects. Indeed the highest 0.2--2.3 keV luminosity found is only $\sim 3\times10^{41}\rm~erg~s^{-1}$, which would correspond to $\sim$1\% Eddington even for a black hole mass as low as $10^5\rm~M_\odot$. 
%Only a small fraction of these galaxies are previously known to host a nuclear X-ray source. 
For comparison, the maximum luminosity of the X-ray nuclei detected by \citet{2017ApJ...835..223S} well exceeds this value, regardless of the host galaxy morphological type.
A potential issue here is that the Virgo and Fornax nuclear X-ray sources are essentially detected in soft X-rays, which might be subject to circumnuclear absorption. However, it is unlikely that a large number of obscured AGNs are present in present-day clusters, in which gas-rich galaxies are not a significant population. 
The few detections of nuclear sources in the eRASS1 2.3--5 keV catalog supports this notion. 
Alternatively, as found by \citet{2020MNRAS.492.2268B, 2022MNRAS.510.4556B}, the steep X-ray AGN luminosity function naturally implies a rarity of luminous objects in volume-limited samples.
Regardless, the first complete census of AGNs in Virgo and Fornax in the soft X-rays would provide useful constraints for cosmological simulations of galaxy cluster formation \citep{2023arXiv231106338N}.

A small but non-negligible fraction of the eRASS1 nuclear X-ray sources are associated with dwarf galaxies, which may host a dwarf MBH. 
High-resolution {\it Chandra} observations are desired to determine whether these sources are genuine AGNs. 
Moveover, some of the eRASS1 nuclear X-ray sources have been previously detected or undetected by {\it Chandra} observations (Section~\ref{sec:nucleus}). The subsequent scans of eROSITA would shed light on the flux variability of these sources.

%UCDs not likely

%Alternatively, full eROSITA survey to probe outburst/TDE (EP/CSST)

\begin{acknowledgments}
M.H. is supported by the National Natural Science Foundation of China (grant 12203001) and the fellowship of China National Postdoctoral Program for Innovation Talents (grant BX2021016). 
Z.H. and Z.L. acknowledge the support of the National Natural Science Foundation of China (grant 12225302) and the National Key Research and Development Program of China (NO.2022YFF0503402).
The authors wish to thank Lin He for help with the eRASS1 data. 

This work is based on data from eROSITA, the soft X-ray instrument aboard SRG, a joint Russian-German science mission supported by the Russian Space Agency (Roskosmos), in the interests of the Russian Academy of Sciences represented by its Space Research Institute (IKI), and the Deutsches Zentrum f\"{u}r Luftund Raumfahrt (DLR). The SRG spacecraft was built by Lavochkin Association (NPOL) and its subcontractors, and is operated by NPOL with support from the Max Planck Institute for Extraterrestrial Physics (MPE). The development and construction of the eROSITA X-ray instrument was led by MPE, with contributions from the Dr. Karl Remeis Observatory Bamberg \& ECAP (FAU Erlangen-Nuernberg), the University of Hamburg Observatory, the Leibniz Institute for Astrophysics Potsdam (AIP), and the Institute for Astronomy and Astrophysics of the University of T\"{u}bingen, with the support of DLR and the Max Planck Society. The Argelander Institute for Astronomy of the University of Bonn and the Ludwig Maximilians Universit\"{a}t Munich also participated in the science preparation for eROSITA.
\end{acknowledgments}

%\begin{appendix}

%\begin{figure*}[ht!]
%\caption{eROSITA 0.2--2.3 keV images and DESI ??-band images of the Virgo galaxies that host nuclear X-ray sources. The name of each galaxy is adopted from the EVCC. All panels have a size of ??.}  \label{fig:individual-Virgo}
%\end{figure*}

%\begin{figure*}[ht!]
%\caption{eROSITA 0.2--2.3 keV images and DSS $B$-band images of the Fornax galaxies that host nuclear X-ray sources. The name of each galaxy is adopted from the FCC. All panels have a size of ??.} 
% \label{fig:individual-Fornax}
%\end{figure*}

%\end{appendix}

\bibliography{eRASS1_AGN_manuscript}

\begin{thebibliography}{}
\expandafter\ifx\csname natexlab\endcsname\relax\def\natexlab#1{#1}\fi
\providecommand{\url}[1]{\href{#1}{#1}}
\providecommand{\dodoi}[1]{doi:~\href{http://doi.org/#1}{\nolinkurl{#1}}}
\providecommand{\doeprint}[1]{\href{http://ascl.net/#1}{\nolinkurl{http://ascl.net/#1}}}
\providecommand{\doarXiv}[1]{\href{https://arxiv.org/abs/#1}{\nolinkurl{https://arxiv.org/abs/#1}}}

\bibitem[{{Bell} {et~al.}(2003){Bell}, {McIntosh}, {Katz}, \& {Weinberg}}]{2003ApJS..149..289B}
{Bell}, E.~F., {McIntosh}, D.~H., {Katz}, N., \& {Weinberg}, M.~D. 2003, \apjs, 149, 289, \dodoi{10.1086/378847}

\bibitem[{{Binggeli} {et~al.}(1985){Binggeli}, {Sandage}, \& {Tammann}}]{1985AJ.....90.1681B}
{Binggeli}, B., {Sandage}, A., \& {Tammann}, G.~A. 1985, \aj, 90, 1681, \dodoi{10.1086/113874}

\bibitem[{{Birchall} {et~al.}(2020){Birchall}, {Watson}, \& {Aird}}]{2020MNRAS.492.2268B}
{Birchall}, K.~L., {Watson}, M.~G., \& {Aird}, J. 2020, \mnras, 492, 2268, \dodoi{10.1093/mnras/staa040}

\bibitem[{{Birchall} {et~al.}(2022){Birchall}, {Watson}, {Aird}, \& {Starling}}]{2022MNRAS.510.4556B}
{Birchall}, K.~L., {Watson}, M.~G., {Aird}, J., \& {Starling}, R.~L.~C. 2022, \mnras, 510, 4556, \dodoi{10.1093/mnras/stab3573}

\bibitem[{{Blakeslee} {et~al.}(2009){Blakeslee}, {Jord{\'a}n}, {Mei}, {C{\^o}t{\'e}}, {Ferrarese}, {Infante}, {Peng}, {Tonry}, \& {West}}]{2009ApJ...694..556B}
{Blakeslee}, J.~P., {Jord{\'a}n}, A., {Mei}, S., {et~al.} 2009, \apj, 694, 556, \dodoi{10.1088/0004-637X/694/1/556}

\bibitem[{{Boselli} {et~al.}(2022){Boselli}, {Fossati}, \& {Sun}}]{2022A&ARv..30....3B}
{Boselli}, A., {Fossati}, M., \& {Sun}, M. 2022, \aapr, 30, 3, \dodoi{10.1007/s00159-022-00140-3}

\bibitem[{{Drinkwater} {et~al.}(2001){Drinkwater}, {Gregg}, {Holman}, \& {Brown}}]{2001MNRAS.326.1076D}
{Drinkwater}, M.~J., {Gregg}, M.~D., {Holman}, B.~A., \& {Brown}, M.~J.~I. 2001, \mnras, 326, 1076, \dodoi{10.1046/j.1365-8711.2001.04646.x}

\bibitem[{{Fabian}(2012)}]{2012ARA&A..50..455F}
{Fabian}, A.~C. 2012, \araa, 50, 455, \dodoi{10.1146/annurev-astro-081811-125521}

\bibitem[{{Ferguson}(1989)}]{1989AJ.....98..367F}
{Ferguson}, H.~C. 1989, \aj, 98, 367, \dodoi{10.1086/115152}

\bibitem[{{Gallo} {et~al.}(2008){Gallo}, {Treu}, {Jacob}, {Woo}, {Marshall}, \& {Antonucci}}]{2008ApJ...680..154G}
{Gallo}, E., {Treu}, T., {Jacob}, J., {et~al.} 2008, \apj, 680, 154, \dodoi{10.1086/588012}

\bibitem[{{Gallo} {et~al.}(2010){Gallo}, {Treu}, {Marshall}, {Woo}, {Leipski}, \& {Antonucci}}]{2010ApJ...714...25G}
{Gallo}, E., {Treu}, T., {Marshall}, P.~J., {et~al.} 2010, \apj, 714, 25, \dodoi{10.1088/0004-637X/714/1/25}

\bibitem[{{Giacconi} {et~al.}(2002){Giacconi}, {Zirm}, {Wang}, {Rosati}, {Nonino}, {Tozzi}, {Gilli}, {Mainieri}, {Hasinger}, {Kewley}, {Bergeron}, {Borgani}, {Gilmozzi}, {Grogin}, {Koekemoer}, {Schreier}, {Zheng}, \& {Norman}}]{2002ApJS..139..369G}
{Giacconi}, R., {Zirm}, A., {Wang}, J., {et~al.} 2002, \apjs, 139, 369, \dodoi{10.1086/338927}

\bibitem[{{Goulding} {et~al.}(2023){Goulding}, {Greene}, {Setton}, {Labbe}, {Bezanson}, {Miller}, {Atek}, {Bogd{\'a}n}, {Brammer}, {Chemerynska}, {Cutler}, {Dayal}, {Fudamoto}, {Fujimoto}, {Furtak}, {Kokorev}, {Khullar}, {Leja}, {Marchesini}, {Natarajan}, {Nelson}, {Oesch}, {Pan}, {Papovich}, {Price}, {van Dokkum}, {Wang}, {Weaver}, {Whitaker}, \& {Zitrin}}]{2023ApJ...955L..24G}
{Goulding}, A.~D., {Greene}, J.~E., {Setton}, D.~J., {et~al.} 2023, \apjl, 955, L24, \dodoi{10.3847/2041-8213/acf7c5}

\bibitem[{{Graham} {et~al.}(2021){Graham}, {Soria}, {Davis}, {Kolehmainen}, {Maccarone}, {Miller-Jones}, {Motch}, \& {Swartz}}]{2021ApJ...923..246G}
{Graham}, A.~W., {Soria}, R., {Davis}, B.~L., {et~al.} 2021, \apj, 923, 246, \dodoi{10.3847/1538-4357/ac34f4}

\bibitem[{{Greene} {et~al.}(2020){Greene}, {Strader}, \& {Ho}}]{2020ARA&A..58..257G}
{Greene}, J.~E., {Strader}, J., \& {Ho}, L.~C. 2020, \araa, 58, 257, \dodoi{10.1146/annurev-astro-032620-021835}

\bibitem[{{Ho}(2008)}]{2008ARA&A..46..475H}
{Ho}, L.~C. 2008, \araa, 46, 475, \dodoi{10.1146/annurev.astro.45.051806.110546}

\bibitem[{{Hou} {et~al.}(2024){Hou}, {He}, {Hu}, {Li}, {Jones}, {Forman}, {Su}, {Wang}, \& {Ho}}]{2024ApJ...961..249H}
{Hou}, M., {He}, L., {Hu}, Z., {et~al.} 2024, \apj, 961, 249, \dodoi{10.3847/1538-4357/ad138a}

\bibitem[{{Hu} {et~al.}(2023){Hu}, {Su}, {Li}, {Hess}, {Kraft}, {Forman}, {Nulsen}, {Sridhar}, {Stroe}, {Baek}, {Chung}, {Grupe}, {Chen}, {Irwin}, {Jones}, {Randall}, \& {Roediger}}]{2023ApJ...956..104H}
{Hu}, Z., {Su}, Y., {Li}, Z., {et~al.} 2023, \apj, 956, 104, \dodoi{10.3847/1538-4357/acf292}

\bibitem[{{Jarrett} {et~al.}(2019){Jarrett}, {Cluver}, {Brown}, {Dale}, {Tsai}, \& {Masci}}]{2019ApJS..245...25J}
{Jarrett}, T.~H., {Cluver}, M.~E., {Brown}, M.~J.~I., {et~al.} 2019, \apjs, 245, 25, \dodoi{10.3847/1538-4365/ab521a}

\bibitem[{{Kim} {et~al.}(2014){Kim}, {Rey}, {Jerjen}, {Lisker}, {Sung}, {Lee}, {Chung}, {Pak}, {Yi}, \& {Lee}}]{2014ApJS..215...22K}
{Kim}, S., {Rey}, S.-C., {Jerjen}, H., {et~al.} 2014, \apjs, 215, 22, \dodoi{10.1088/0067-0049/215/2/22}

\bibitem[{{Kormendy} \& {Ho}(2013)}]{2013ARA&A..51..511K}
{Kormendy}, J., \& {Ho}, L.~C. 2013, \araa, 51, 511, \dodoi{10.1146/annurev-astro-082708-101811}

\bibitem[{{Lee} {et~al.}(2019){Lee}, {Gallo}, {Hodges-Kluck}, {Cot{\'e}}, {Ferrarese}, {Miller}, {Baldassare}, {Plotkin}, \& {Treu}}]{2019ApJ...874...77L}
{Lee}, N., {Gallo}, E., {Hodges-Kluck}, E., {et~al.} 2019, \apj, 874, 77, \dodoi{10.3847/1538-4357/ab05cd}

\bibitem[{{Lehmer} {et~al.}(2010){Lehmer}, {Alexander}, {Bauer}, {Brandt}, {Goulding}, {Jenkins}, {Ptak}, \& {Roberts}}]{2010ApJ...724..559L}
{Lehmer}, B.~D., {Alexander}, D.~M., {Bauer}, F.~E., {et~al.} 2010, \apj, 724, 559, \dodoi{10.1088/0004-637X/724/1/559}

\bibitem[{{Lehmer} {et~al.}(2016){Lehmer}, {Basu-Zych}, {Mineo}, {Brandt}, {Eufrasio}, {Fragos}, {Hornschemeier}, {Luo}, {Xue}, {Bauer}, {Gilfanov}, {Ranalli}, {Schneider}, {Shemmer}, {Tozzi}, {Trump}, {Vignali}, {Wang}, {Yukita}, \& {Zezas}}]{2016ApJ...825....7L}
{Lehmer}, B.~D., {Basu-Zych}, A.~R., {Mineo}, S., {et~al.} 2016, \apj, 825, 7, \dodoi{10.3847/0004-637X/825/1/7}

\bibitem[{{Luo} {et~al.}(2017){Luo}, {Brandt}, {Xue}, {Lehmer}, {Alexander}, {Bauer}, {Vito}, {Yang}, {Basu-Zych}, {Comastri}, {Gilli}, {Gu}, {Hornschemeier}, {Koekemoer}, {Liu}, {Mainieri}, {Paolillo}, {Ranalli}, {Rosati}, {Schneider}, {Shemmer}, {Smail}, {Sun}, {Tozzi}, {Vignali}, \& {Wang}}]{2017ApJS..228....2L}
{Luo}, B., {Brandt}, W.~N., {Xue}, Y.~Q., {et~al.} 2017, \apjs, 228, 2, \dodoi{10.3847/1538-4365/228/1/2}

\bibitem[{{Mei} {et~al.}(2007){Mei}, {Blakeslee}, {C{\^o}t{\'e}}, {Tonry}, {West}, {Ferrarese}, {Jord{\'a}n}, {Peng}, {Anthony}, \& {Merritt}}]{2007ApJ...655..144M}
{Mei}, S., {Blakeslee}, J.~P., {C{\^o}t{\'e}}, P., {et~al.} 2007, \apj, 655, 144, \dodoi{10.1086/509598}

\bibitem[{{Merloni} {et~al.}(2024){Merloni}, {Lamer}, {Liu}, {Ramos-Ceja}, {Brunner}, {Bulbul}, {Dennerl}, {Doroshenko}, {Freyberg}, {Friedrich}, {Gatuzz}, {Georgakakis}, {Haberl}, {Igo}, {Kreykenbohm}, {Liu}, {Maitra}, {Malyali}, {Mayer}, {Nandra}, {Predehl}, {Robrade}, {Salvato}, {Sanders}, {Stewart}, {Tub{\'\i}n-Arenas}, {Weber}, {Wilms}, {Arcodia}, {Artis}, {Aschersleben}, {Avakyan}, {Aydar}, {Bahar}, {Balzer}, {Becker}, {Berger}, {Boller}, {Bornemann}, {Br{\"u}ggen}, {Brusa}, {Buchner}, {Burwitz}, {Camilloni}, {Clerc}, {Comparat}, {Coutinho}, {Czesla}, {Dannhauer}, {Dauner}, {Dauser}, {Dietl}, {Dolag}, {Dwelly}, {Egg}, {Ehl}, {Freund}, {Friedrich}, {Gaida}, {Garrel}, {Ghirardini}, {Gokus}, {Gr{\"u}nwald}, {Grandis}, {Grotova}, {Gruen}, {Gueguen}, {H{\"a}mmerich}, {Hamaus}, {Hasinger}, {Haubner}, {Homan}, {Ider Chitham}, {Joseph}, {Joyce}, {K{\"o}nig}, {Kaltenbrunner}, {Khokhriakova}, {Kink}, {Kirsch}, {Kluge}, {Knies}, {Krippendorf}, {Krumpe}, {Kurpas}, {Li}, {Liu}, {Locatelli}, {Lorenz}, {M{\"u}ller},
  {Magaudda}, {Mannes}, {McCall}, {Meidinger}, {Michailidis}, {Migkas}, {Mu{\~n}oz-Giraldo}, {Musiimenta}, {Nguyen-Dang}, {Ni}, {Olechowska}, {Ota}, {Pacaud}, {Pasini}, {Perinati}, {Pires}, {Pommranz}, {Ponti}, {Poppenhaeger}, {P{\"u}hlhofer}, {Rau}, {Reh}, {Reiprich}, {Roster}, {Saeedi}, {Santangelo}, {Sasaki}, {Schmitt}, {Schneider}, {Schrabback}, {Schuster}, {Schwope}, {Seppi}, {Serim}, {Shreeram}, {Sokolova-Lapa}, {Starck}, {Stelzer}, {Stierhof}, {Suleimanov}, {Tenzer}, {Traulsen}, {Tr{\"u}mper}, {Tsuge}, {Urrutia}, {Veronica}, {Waddell}, {Willer}, {Wolf}, {Yeung}, {Zainab}, {Zangrandi}, {Zhang}, {Zhang}, \& {Zheng}}]{2024A&A...682A..34M}
{Merloni}, A., {Lamer}, G., {Liu}, T., {et~al.} 2024, \aap, 682, A34, \dodoi{10.1051/0004-6361/202347165}

\bibitem[{{Miller} {et~al.}(2012){Miller}, {Gallo}, {Treu}, \& {Woo}}]{2012ApJ...747...57M}
{Miller}, B., {Gallo}, E., {Treu}, T., \& {Woo}, J.-H. 2012, \apj, 747, 57, \dodoi{10.1088/0004-637X/747/1/57}

\bibitem[{{Nelson} {et~al.}(2023){Nelson}, {Pillepich}, {Ayromlou}, {Lee}, {Lehle}, {Rohr}, \& {Truong}}]{2023arXiv231106338N}
{Nelson}, D., {Pillepich}, A., {Ayromlou}, M., {et~al.} 2023, arXiv e-prints, arXiv:2311.06338, \dodoi{10.48550/arXiv.2311.06338}

\bibitem[{{Padovani} {et~al.}(2017){Padovani}, {Alexander}, {Assef}, {De Marco}, {Giommi}, {Hickox}, {Richards}, {Smol{\v{c}}i{\'c}}, {Hatziminaoglou}, {Mainieri}, \& {Salvato}}]{2017A&ARv..25....2P}
{Padovani}, P., {Alexander}, D.~M., {Assef}, R.~J., {et~al.} 2017, \aapr, 25, 2, \dodoi{10.1007/s00159-017-0102-9}

\bibitem[{{Schmidt}(1963)}]{1963Natur.197.1040S}
{Schmidt}, M. 1963, \nat, 197, 1040, \dodoi{10.1038/1971040a0}

\bibitem[{{She} {et~al.}(2017){She}, {Ho}, \& {Feng}}]{2017ApJ...835..223S}
{She}, R., {Ho}, L.~C., \& {Feng}, H. 2017, \apj, 835, 223, \dodoi{10.3847/1538-4357/835/2/223}

\bibitem[{{Shi} {et~al.}(2021){Shi}, {Li}, {Yuan}, \& {Zhu}}]{2021NatAs...5..928S}
{Shi}, F., {Li}, Z., {Yuan}, F., \& {Zhu}, B. 2021, Nature Astronomy, 5, 928, \dodoi{10.1038/s41550-021-01394-0}

\bibitem[{{Simionescu} {et~al.}(2017){Simionescu}, {Werner}, {Mantz}, {Allen}, \& {Urban}}]{2017MNRAS.469.1476S}
{Simionescu}, A., {Werner}, N., {Mantz}, A., {Allen}, S.~W., \& {Urban}, O. 2017, \mnras, 469, 1476, \dodoi{10.1093/mnras/stx919}

\bibitem[{{Soria} {et~al.}(2022){Soria}, {Kolehmainen}, {Graham}, {Swartz}, {Yukita}, {Motch}, {Jarrett}, {Miller-Jones}, {Plotkin}, {Maccarone}, {Ferrarese}, {Guest}, \& {Lan{\c{c}}on}}]{2022MNRAS.512.3284S}
{Soria}, R., {Kolehmainen}, M., {Graham}, A.~W., {et~al.} 2022, \mnras, 512, 3284, \dodoi{10.1093/mnras/stac148}

\bibitem[{{Sunyaev} {et~al.}(2021){Sunyaev}, {Arefiev}, {Babyshkin}, {Bogomolov}, {Borisov}, {Buntov}, {Brunner}, {Burenin}, {Churazov}, {Coutinho}, {Eder}, {Eismont}, {Freyberg}, {Gilfanov}, {Gureyev}, {Hasinger}, {Khabibullin}, {Kolmykov}, {Komovkin}, {Krivonos}, {Lapshov}, {Levin}, {Lomakin}, {Lutovinov}, {Medvedev}, {Merloni}, {Mernik}, {Mikhailov}, {Molodtsov}, {Mzhelsky}, {M{\"u}ller}, {Nandra}, {Nazarov}, {Pavlinsky}, {Poghodin}, {Predehl}, {Robrade}, {Sazonov}, {Scheuerle}, {Shirshakov}, {Tkachenko}, \& {Voron}}]{2021A&A...656A.132S}
{Sunyaev}, R., {Arefiev}, V., {Babyshkin}, V., {et~al.} 2021, \aap, 656, A132, \dodoi{10.1051/0004-6361/202141179}

\bibitem[{{Weinberger} {et~al.}(2017){Weinberger}, {Springel}, {Hernquist}, {Pillepich}, {Marinacci}, {Pakmor}, {Nelson}, {Genel}, {Vogelsberger}, {Naiman}, \& {Torrey}}]{2017MNRAS.465.3291W}
{Weinberger}, R., {Springel}, V., {Hernquist}, L., {et~al.} 2017, \mnras, 465, 3291, \dodoi{10.1093/mnras/stw2944}

\bibitem[{{Wright} {et~al.}(2010){Wright}, {Eisenhardt}, {Mainzer}, {Ressler}, {Cutri}, {Jarrett}, {Kirkpatrick}, {Padgett}, {McMillan}, {Skrutskie}, {Stanford}, {Cohen}, {Walker}, {Mather}, {Leisawitz}, {Gautier}, {McLean}, {Benford}, {Lonsdale}, {Blain}, {Mendez}, {Irace}, {Duval}, {Liu}, {Royer}, {Heinrichsen}, {Howard}, {Shannon}, {Kendall}, {Walsh}, {Larsen}, {Cardon}, {Schick}, {Schwalm}, {Abid}, {Fabinsky}, {Naes}, \& {Tsai}}]{2010AJ....140.1868W}
{Wright}, E.~L., {Eisenhardt}, P. R.~M., {Mainzer}, A.~K., {et~al.} 2010, \aj, 140, 1868, \dodoi{10.1088/0004-6256/140/6/1868}

\bibitem[{{Xue}(2017)}]{2017NewAR..79...59X}
{Xue}, Y.~Q. 2017, \nar, 79, 59, \dodoi{10.1016/j.newar.2017.09.002}

\bibitem[{{Yoon} {et~al.}(2019){Yoon}, {Yuan}, {Ostriker}, {Ciotti}, \& {Zhu}}]{2019ApJ...885...16Y}
{Yoon}, D., {Yuan}, F., {Ostriker}, J.~P., {Ciotti}, L., \& {Zhu}, B. 2019, \apj, 885, 16, \dodoi{10.3847/1538-4357/ab45e8}

\bibitem[{{Yuan} \& {Narayan}(2014)}]{2014ARA&A..52..529Y}
{Yuan}, F., \& {Narayan}, R. 2014, \araa, 52, 529, \dodoi{10.1146/annurev-astro-082812-141003}

\end{thebibliography}
\bibliographystyle{aasjournal}

\end{document}